\documentstyle[prl,aps,epsf,twocolumn]{revtex}
\title
{Mixed Heisenberg Chains. I. The Ground State Problem.}
\author{Harald Niggemann, Gennadi Uimin\cite{gu}, and Johannes Zittartz}
\address{Institut f\"ur Theoretische Physik, 
Universit\"at zu K\"oln, Z\"ulpicher Str.77, 
D-50937 K\"oln, Germany}
\address{
{\small (Received ... March 1997)}}
\address{\vspace{10pt}}
\author{\small\parbox{14.1cm}{\small
We consider a mechanism for competing interactions in alternating
Heisenberg spin chains due to the formation of local spin-singlet pairs.
The competition of spin-1 and spin-0 states reveals hidden Ising symmetry of
such alternating chains. 
\\
\\
PACS numbers:}}
\begin{document}
\maketitle
\subsection*{Introduction}
During the last few years mixed quantum spin chains have attracted
some interest of theorists. Exactly solvable versions with 
sophisticated Hamiltonians
have been studied via Bethe ansatz \cite{vega,fujii,doerfel}. Very recently 
numerical methods \cite{pati,brehmer} and matrix-product
techniques \cite{kolezhuk} have been applied to these spin systems.
Different kinds of alternating chains with XXZ-like interactions have been
investigated by using finite-size calculations and conformal invariance \cite{alcaraz}.
The subject of interest in this paper are alternating spin chains, in which
each second site of the chain is considered as a compound, 
a kind of dumbbell configuration.
Two dumbbell spins 1/2 interact with each other via the Heisenberg interaction
with a coupling constant $J_0$, either ferromagnetically, 
or antiferromagnetically. 
Each first site of the chain is occupied (A) by a usual quantum spin 
(we shall consider the cases of spin $1/2,\, 1,\, 3/2,\, 2$), or
(B) by a compound spin too (see Fig. 1).
This spin is supposed to be antiferromagnetically
coupled (coupling constant $J_1<0$) to the spins of nearest dumbbells.
In spite of short range Heisenberg interactions, variations
of $J_0$ may result in first order transitions at zero temperature.

Realizations of such one-dimensional chains 
are shown in Fig. 1(a and b).
The Hamiltonian in case A can be written as
\begin{eqnarray}
\label{ham_a}
H^{(a)}=-{\cal J}_1\sum_{<\!\rho,r\!>}{\bf s}(\rho)\cdot
({\bbox\sigma}(r_1)+{\bbox\sigma}(r_2))\nonumber\\
-{\cal J}_0\sum_{<\!r_1,r_2\!>}
{\bbox\sigma}(r_1)\cdot{\bbox\sigma}(r_2),
\end{eqnarray}
whereas in case B it becomes
\begin{eqnarray}
\label{ham_b}
H^{(b)}=-{\cal J}_1\sum_{<\!\rho,r\!>}({\bf s}(\rho_1)+{\bf s}(\rho_2))\cdot
({\bbox\sigma}(r_1)+{\bbox\sigma}(r_2))\nonumber\\
-{\cal J}_0'\sum_{<\!\rho_1,\rho_2\!>}
{\bf s}(\rho_1)\cdot{\bf s}(\rho_2)-
{\cal J}_0\sum_{<\!r_1,r_2\!>}
{\bbox\sigma}(r_1)\cdot{\bbox\sigma}(r_2).
\end{eqnarray}
Except for one special case which is
described by model B, we concentrate our efforts on model A.
The Hamiltonians in (\ref{ham_a}-\ref{ham_b})
can also be represented as follows:
\begin{eqnarray}
\label{ham_1}
H^{(a,b)}=H_1+H_0^{(a,b)},\quad
H_1=-{\cal J}_1\sum_{<\!\rho,r\!>}{\bf S}(\rho)\cdot{\bf S}(r),
\end{eqnarray} 
and
\begin{eqnarray}
\label{ham_a0}
H^{(a)}_0=-\frac 12{\cal J}_0\sum_r{\bf S}^2(r),\\
H^{(b)}_0=-\frac 12{\cal J}_0'\sum_{\rho}{\bf S}^2(\rho)
-\frac 12{\cal J}_0\sum_r{\bf S}^2(r).
\label{ham_b0}
\end{eqnarray}
\vspace{-0.2cm}

\noindent
When making transformations from Hamiltonians (\ref{ham_a}-\ref{ham_b})
to (\ref{ham_a0}-\ref{ham_b0}), we have skipped irrelevant constant terms.
Co-ordinates of spins in dumbbells, $r_1$ and $r_2$, are replaced 
by a single co-ordinate $r$ (in model B $\rho_1$ and $\rho_2$ are also
replaced by their common $\rho$).
Note that model B transforms into model A when 
${\cal J}_0'\to\-\infty.$
\begin{figure}
\epsfxsize=70mm
\epsffile{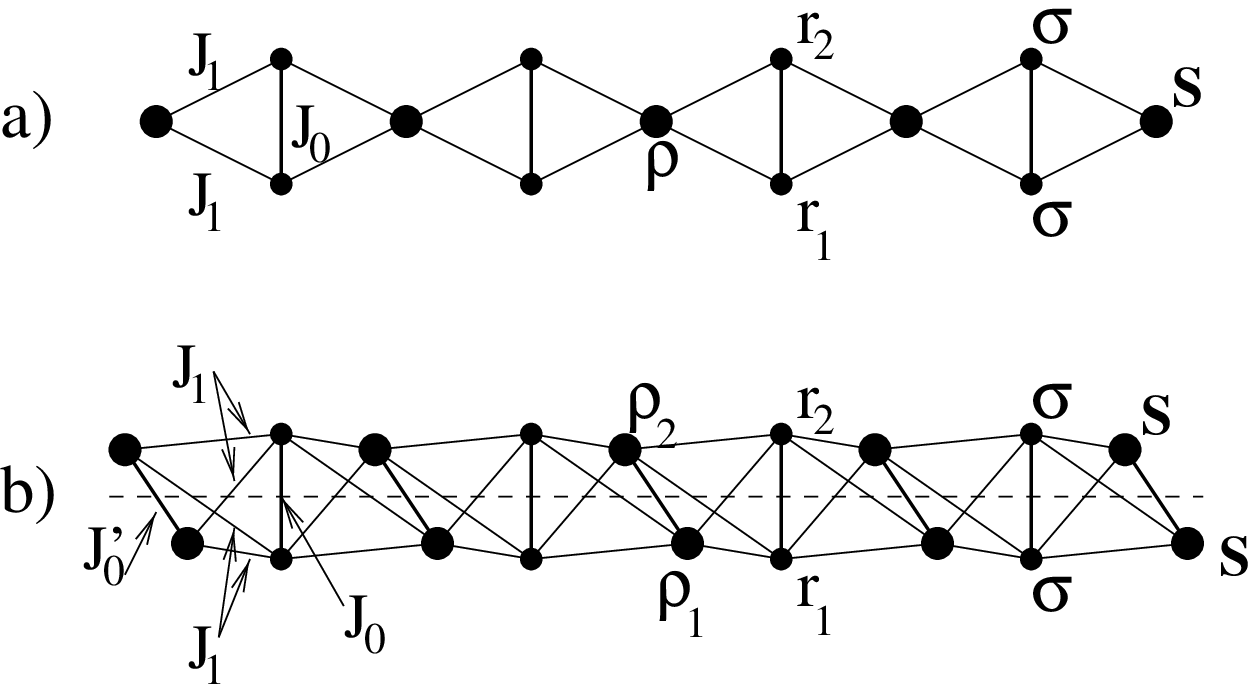}
\hfill

\noindent
\label{figure1}
\baselineskip=10pt
{{\small FIG.\ 1\quad  Shown are chain fragments of length $k=4$.
Model $A$: Spins $s$ ($=1/2,\, 1,\, 3/2,\, 2$) occupy sites $\rho$,
spins $\sigma$ are arranged in dumbbells. Model $B$: Spins $s$ are in
dumbbells, too, which are shown orthogonal to the $\sigma$-dumbbells.
}}
\end{figure}
The two spins, ${\bbox\sigma}(r_1)$ and ${\bbox\sigma}(r_2)$,
are incorporated into the {\it compound} spin: ${\bf S}(r)=
{\bbox\sigma}(r_1)+{\bbox\sigma}(r_2)$, which is either 0, or 1.
This reveals a hidden Ising symmetry of the original Heisenberg models 
(\ref{ham_a}) and (\ref{ham_b}).
In fact, Hamiltonian $H_1$ does not generate any transitions between 
the total spin states 0 and 1 of any compound spin. 
Thus we can introduce spin-0 states on (some) $r$-sites
which are kind of intrinsic ``defects''. Governed by the ${\cal J}_0$-terms,
these ``defects'' regulate a separation of the original chain into an ensemble
of finite chain fragments decoupled from each other. 
Their structure can be defined as follows:
A chain fragment of length $k$ ($k\geq 1$) consists of $k+1$ spins $s$ and 
$k$ spins 1. The spins of these two groups alternate with each other.
A chain fragment can be formally described as 
$(s, 1)^ks$. 
Chain fragments are isolated from each other by spins 0. 

For convenience, we enumerate the lattice sites $r$ 
by integer numbers, then half-integers are reserved for sites $\rho$.

All energies are suitably measured in units of ${\cal J}_1$ which is
supposed to be negative. Thus, we set ${\cal J}_1=-1$.
Below we shall determine the phase diagram of the chains with 
$s=1/2,\,1,\,3/2,\,2$ depending on the parameter ${\cal J}_0$ in model A,
and both, ${\cal J}_0$ and ${\cal J}_0'$, in model B.

Our analysis includes elements of a rigorous analytical approach,
a linear programming method, and numerical methods.
Here we use the density matrix renormalization
group (DMRG) method \cite{white1,white2} which is most appropriate to our
problem.

A detailed description of the DMRG algorithm which we have used
to compute ground state energies of finite open chains can
be found in Appendix A. It was implemented in C++ and ran on
a SUN UltraSparc 2 workstation with two 167 MHz processors and
256 MB memory. In order to achieve the desired accuracy we
kept up to $N=100$ block states during each DMRG step. The
whole project consumed about 400 hours of CPU time. No
parallelization was used.
\subsection*{The ground state problem. Model A}
For convenience, we include into the definition of a chain fragment of 
length $k$ a spin-0 state, say, from its
right. Then a chain fragment of length $k$ is represented
by $(s, 1)^k(s,0)$.
This classification needs the ``empty'' chain fragments to be included:
Any of these ``empty'' fragments is spin $s$ with spin 0, attached from 
its right, i.e. $(s, 0)$. Conventionally, a nearest 
spin from the left of any chain fragment {\it is also} 0, but it is 
incorporated into the nearest-from-left chain fragment.

Let us suppose that we have succeeded to determine the ground state energies
of Hamiltonian $H_1$ for all finite chain fragments, i.e. $\{\epsilon_0,
\epsilon_1, \epsilon_2,\dots,\epsilon_k,\dots\}$. Then, the contribution of
$H_0^{(a)}$ will be $- k{\cal J}_0$ for the chain fragment of length $k$.
The ground state energy (per compound spin) of the chain, consisting of 
$N_0$ "empty" chain fragments, $N_1$ chain fragments of length 1, \ldots, 
$N_k$ chain fragments of length $k$, etc. is
\begin{eqnarray}
\label{gs_en}
{\cal E_{\rm g.s}}=\sum_{k\geq 0}(\epsilon_k - k{\cal J}_0)w_k.
\end{eqnarray}
In (\ref{gs_en}) we introduced ``probabilities'' $w_k=N_k/{\cal N}, k\geq 0$ 
and ${\cal N}$ being the total number of compound spins.
Expressing ${\cal N}$ in terms of the numbers $N_k$ of the
various chain fragments is equivalent to the constraint 
\begin{eqnarray}
\label{sites}
1=\sum_{k\geq 0}(k+1)w_k
\end{eqnarray}
imposed on the set of ``probabilities''.
Of course, all $w_k$ are non-negative.
Eqs. (\ref{gs_en})-(\ref{sites}) constitute a linear programming problem 
which demands
that extrema of the energy (\ref{gs_en}) should be searched at the vertices of 
the polygon defined by (\ref{sites}). This method has been efficiently
applied to the problem of competative interactions, leading to complex
modulated structures \cite{hubb,u_p1,f_s,u_p2}.
For this particular problem, it can be easily proven that each vertex 
is characterized by only one non-zero ``probability'' value.
For example, the vertex with $w_0=1$ corresponds to the perfect 
structure with the periodicity element $(s, 0)$, 
its energy being ${\cal E}=0$ (we denote this regular spin configuration 
by $\langle 0\rangle $); 
$w_1=1/2$ (a configuration conventionally denoted by $\langle 1\rangle $)
corresponds to the periodicity element $(s,1,s,0)$
with energy ${\cal E}=(\epsilon_1 - {\cal J}_0)/2$.
For the $\langle k\rangle $ state, $N_k={\cal N}/(k+1)$ and 
the periodicity element can be represented as ($(s,1)^k,s,0$).
The energy of this spin configuration is given by 
\begin{equation}
\label{g.s.ener}
{\cal E}_k=(\epsilon_k - k{\cal J}_0)/(k+1).
\end{equation}

Numerical methods which are outlined in Appendix A allow us to analyze
characteristic first order transitions. They happen at zero temperature
and are controlled by ${\cal J}_0$. For $s=1/2$ the set of energies 
$\{\epsilon_i\}$ is given in Table I (left column). 
According to the Lieb-Mattis theorem, 
a ferrimagnetic ground state 
would be realized with a total spin $k/2$, if we dealt with a periodic
alternating chain, consisting of $k$ spins 1/2 and $k$ spins 1, coupled
antiferromagnetically. In our case,
a chain fragment with $k$ spins 1 and $k+1$ spins 1/2 will exhibit the total 
spin $S_k=(k-1)/2 \;(k\geq 1)$ in the ground state. 
Low lying excitations are asymmetric.
Their hierarchy is as follows: the lowest excitations are triplets which
correspond to $\Delta S=-1$, or the total spin $(k-3)/2 \;(k\geq 3)$. The next
lowest excitations are triplets too, but with $\Delta S=+1$ (the total spin 
$(k+1)/2$). The singlet excitations are lying above both triplet excitations.
Only the triplet excitations with $\Delta S=-1$ will give rise to a gapless 
mode in the limit $k\to\infty$. 
\vspace{-2.5cm}

\begin{figure}
\epsfxsize=75mm
\epsffile{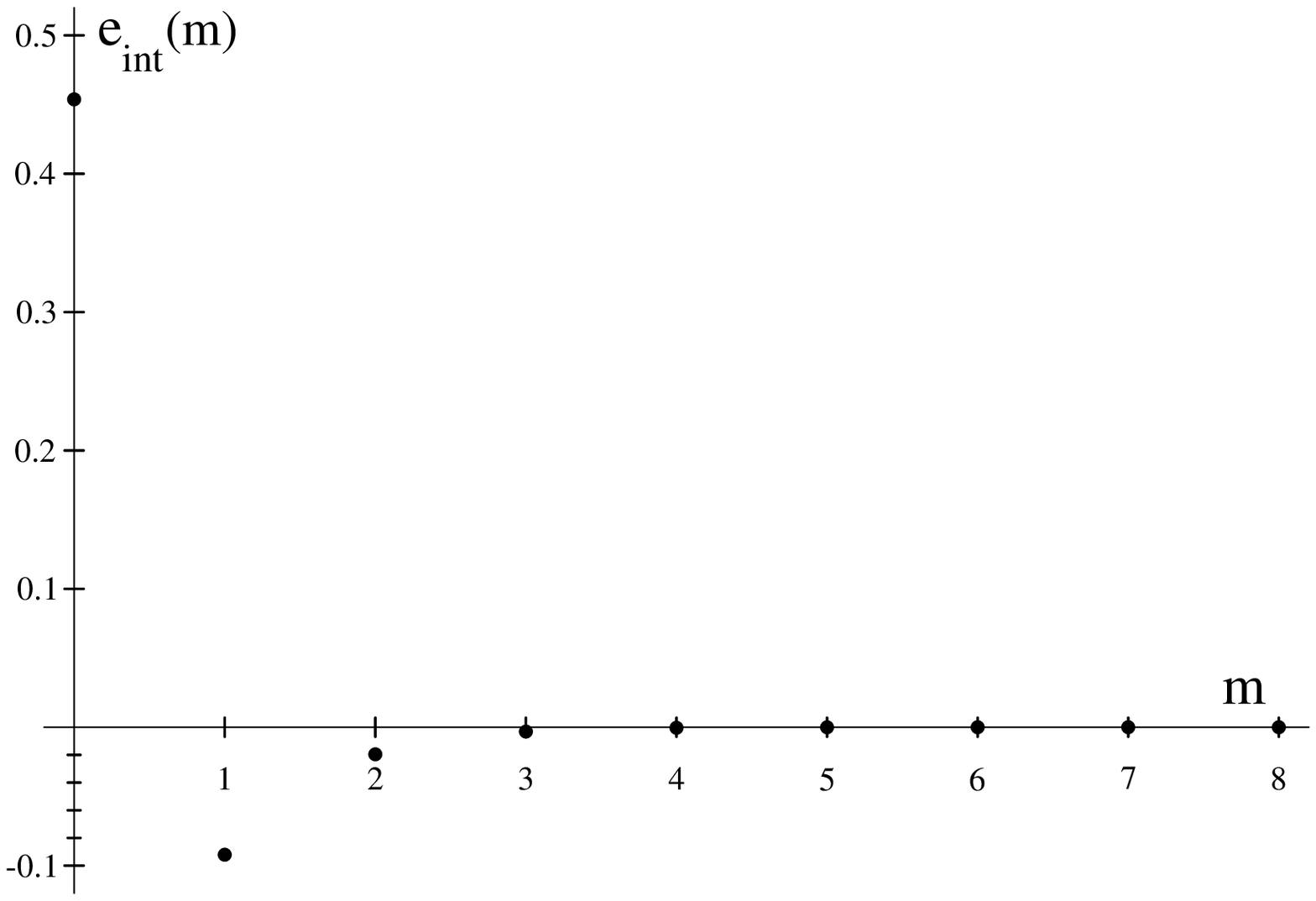}
\vspace{-2.4cm}

\noindent
\label{figure2}
\baselineskip=10pt
{{\small FIG.\ 2\quad  Shown are a few first points of $e_{\rm int}(m)$ 
(model A, $s=1/2$). 
}}
\end{figure}
\vspace{-.2cm}

Comparison of the ground state energy expressions (\ref{g.s.ener}) at various 
vertices clearly shows that three configurations are competitive in 
the global ground state:
This is $\langle 0\rangle $ for 
${\cal J}_0<\!-2$, which changes to $\langle 1\rangle $
for $-2<\!{\cal J}_0<\! 2e_{\infty}+2\approx-0.910$. Beyond this region,
i.e. for $-0.910 <\!{\cal J}_0$, the $\langle \infty\rangle $ state becomes energetically 
favourable.
The numerical data for the set $\{\epsilon_k\}$ can be represented as
\begin{equation}
\label{ener_fit}
\epsilon_k=ke_{\infty}+e_0+e_{\rm int}(k), 
\end{equation}
where
$e_{\infty}\approx-1.45412$ is the energy per element ($s,1,s$) of the perfect
periodic spin structure $\langle \infty\rangle \equiv(s,1)^{\infty}$, $s=1/2$. 
The energy due to the open ends is $e_0\approx-0.45352$, 
and the remaining part,
$e_{\rm int}(k)$, can be interpreted as the interaction between the chain 
fragment
ends which apparently turns to zero for $k\to\infty$. This function 
$e_{\rm int}(k)$
plays an important role in establishing the succession of phase transitions.
In Appendix B we perform a rigorous analysis, according to which 
{\it a succession is given by a broken line, which is concave upwards and 
envelops $e_{\rm int}(k)$ from below.} 
This broken line includes the points corresponding to periodic 
structures with shortest chain fragment ($k=0$ in model A) 
and infinitely long chain fragments, $k\to\infty$.
\vspace{0.3cm}

\begin{table}[b]
\squeezetable
\begin{tabular} {c|lr|lr|lr}
\multicolumn{1}{c|}{\mbox{}} & \multicolumn{2}{c|}{$s=1/2$} &
\multicolumn{2}{c|}{$s=3/2$} & \multicolumn{2}{c}{$s=2$}\\
\hline
$k$ &  $\epsilon_k$ & $e_{\rm int}(k)$ & 
$\epsilon_k$ & $e_{\rm int}(k)$ & $\epsilon_k$ & $e_{\rm int}(k)$\\
\hline
0 & 0  & 0.453733  &      0 & 0.150038 &      0 & 0.111298 \\
1 &  -2  & -0.092173  &     -4 & 0.011951 & -5 & 0.005167 \\
2 & -3.3815016  & -0.019581 &   -7.8727492 & 0.001114 & -9.8987531 & 
0.000282 \\
3 & -4.8191715  & -0.003157 &    -11.735772 & 0.000003 & -14.792914 & 
-0.000009 \\
4 & -6.2705355  & -0.000426 &    -15.597742 & -0.000054 & -19.686790 & 
-0.000017 \\
5 & -7.7242539  & -0.000051 &    -19.459627 & -0.000026 & -24.580654 & 
-0.000012 \\
6 & -9.1783024  & -0.000005 &    -23.321522 & -0.000009 & -29.474518 & 
-0.000007 \\
7 & -10.632391  & 0.000000 &    -27.183429 & -0.000003 & -34.368384 & 
-0.000004 \\
8 & -12.086485  & 0.000000 &    -31.045338 & 0.000000 & -39.262251 & 
-0.000003 \\
9 & -13.540579  & 0.000000 &    -34.907251 & 0.000000 & -44.156119 & 
-0.000002 \\
10 & -14.994673  & 0.000000 &    -38.769163 & 0.000000 & -49.049987  & 
-0.000001
\end{tabular}
\vspace{0.3cm}

\label{table1}
\baselineskip=10pt
{{\small TABLE\ 1\quad  The ground state energies of $H_1$ and the function
$e_{\rm int}(k)$
for the chain fragments, consisting of $k$ spins 1 and $k+1$ spins $s$,
1/2, 3/2 and 2 in three successive columns.}}
\vspace{0.6cm}

\begin{tabular} {c|ll|lr}
$k$ & $\epsilon_{2k}$ & $e_{\rm int}(2k)$ & $\epsilon_{2k+1}$ & 
$e_{\rm int}(2k\!+\!1)$\\
\mbox{} & \mbox{} & \mbox{} & or $\epsilon_k$ & or $e_{\rm int}(k)$\\
\hline
0 &      0    & -1.208   &    0  &  0.193484 \\
1 & -2   &  -0.405032  &           -3  & -0.003548 \\
2 & -4.6457513 & -0.247815   & -5.8302125 & -0.030792 \\
3 & -7.3702750 & -0.169371 & -8.6345320 & -0.032144\\
4 & -10.124637 & -0.120765 & -11.432932 & -0.027575\\
5 & -12.894560 & -0.087720 & -14.230359 & -0.022035\\
6 & -15.674010 & -0.064202  & -17.028266 & -0.016973\\ 
7 & -18.459853 & -0.047076 & -19.827036 & -0.012775\\
8 & -21.250218 & -0.034473 & -22.626683 & -0.009455\\
9 & -24.043879 & -0.025167 & -25.427099 & -0.006902\\
10 & -26.839978 & -0.018298 & -28.228140 & -0.004975\\
11 & -29.637889 & -0.013240 & -31.029675 & -0.003542\\
12 & -32.437147 & -0.009530 & -33.831576 & -0.002474\\
13 & -35.237402 & -0.006818 & -36.633771 & -0.001701\\
14 & -38.038394 & -0.004842 & -39.436175 & -0.001138\\
15 & -40.839927 & -0.003407 & -42.238735 & -0.000730
\end{tabular}
\vspace{0.3cm}

\label{table2}
\baselineskip=10pt
{{\small TABLE\ 2\quad  The ground state energies of $H_1$
and the function $e_{\rm int}(k)$
for the chain fragments, consisting of spins 1. For model B, 
the data of computations are
given in the left (right) column for chain fragments of even (odd)
total lengths, $2k$ ($2k+1$). For model A, only the numerical data of the right
column should be taken.
}}
\end{table}
For $s=1/2$, $e_{\rm int}(k)$ is shown in Fig. 2. Thus, the energetically 
favourable configurations can be $\langle 0\rangle $, 
$\langle 1\rangle $ and $\langle \infty\rangle $.

For $s=1$, the function $e_{\rm int}(k)$ is the upper curve in Fig. 3 
related to odd integers $m=2k+1$. According to Appendix B we have
successive phase transitions

\noindent
$\langle 0\rangle \rightarrow \langle 1\rangle \rightarrow \langle 2\rangle
\rightarrow \langle 3\rangle \rightarrow \langle \infty\rangle $

\noindent
at ${\cal J}_0^{(0,1)}=-3$, ${\cal J}_0^{(1,2)}=-2.660425$, 
${\cal J}_0^{(2,3)}=-2.582746$, ${\cal J}_0^{(3,\infty)}=-2.577340$,
which are determined from Eq. (\ref{g.s.ener}) and data of Table 2
(right column). 
In accordance with the Lieb-Mattis theorem, the value of the spin of the
ground state of a chain fragment is $S_k=1$, independent of length.

For $s=3/2$, the total spin in the ground state is $S_k=(3+k)/2$, whereas for
$s=2$, $S_k=k+2$.
A succession of phase transitions can be easily identified by making use of
Table 1 (middle and right columns). It is the same in these two cases:
\newline 
$\langle 0\rangle \rightarrow\langle 1\rangle \rightarrow\langle 2\rangle 
\rightarrow\langle 3\rangle \rightarrow\langle 4\rangle \rightarrow\langle \infty\rangle $. 
\newline
The last two transitions
occur at ${\cal J}_0^{(3,4)}$ and ${\cal J}_0^{(4,\infty)}$ whose values 
slightly differ from ${\cal J}_0^{(2,3)}$.

Note that a transformation $\langle 0\rangle \rightarrow\langle \infty\rangle $
via a few intermediate first order transitions proceeds due to
quantum effects. In fact, if $H_1$ is confined to a pure Ising form,
then $e_{\rm int}(k)\equiv 0$. The system undergoes a direct transition
$\langle 0\rangle \rightarrow \langle \infty\rangle $ at ${\cal J}_0=-2s$. 
A transition point is highly degenerate: Chain fragments of any length 
are allowed as well as any sequence of them.

In contrary to this, a spin-wave approach ``overestimates'' quantum effects
for $s\geq 3/2$:
$e_{\rm int}(k)$ of the spin-wave approach may not differ much from the 
numerical data,
but for $s\geq 3/2$ it always decays monotonically with $k$ and thus would 
lead always to an infinite
set of first order transitions
$\quad \langle 0\rangle \rightarrow\langle 1\rangle \rightarrow\dots\rightarrow
\langle k\rangle \rightarrow\dots\rightarrow\langle \infty\rangle $.
\newline
Exceptional is the case $s=1/2$ for which the spin-wave approach exhibits 
a global minimum of $e_{\rm int}(k)$ at $k=1$. This results in the same 
succession of transitions $\langle 0\rangle \rightarrow\langle 1\rangle
\rightarrow\langle \infty\rangle $, as obtained within the exact numerical
scheme. 
\subsection*{The ground state problem. Model B}
 
The subject of this Section is model B, where $\sigma$ and $s$
are both spins 1/2, thus the equivalent model described by Hamiltonian
(\ref{ham_1}) deals with chain fragments consisting of 
compound spins 1 only. The ground state energies and excitations of finite 
spin-1 chains described by Hamiltonian $H_1$ have been studied by T.Kennedy
\cite{kennedy} in the framework of the Lanczos method.

It is more suitable to enumerate the chain
by integers, say {\it odd} and {\it even} for $\rho$ and $r$ sites,
respectively.
Our consideration can be restricted to ${\cal J}_0'\geq {\cal J}_0$.
Relation (\ref{ener_fit}), which can be used for spin-1 chain fragments as
well, is apparently not identical for chain fragments consisting of even or 
odd number of sites. 
This does not concern
$e_{\infty}$, whose value is common for both types of chain fragments, 
but concerns $e_0$ \cite{comment} and 
$e_{\rm int}(k)$. 
The total spin in the ground state is zero or one, it depends on
the length of chain fragments, even or odd.

The linear programming method will be used again to select those chain 
fragments which may be candidates to form the ground state.
The following four types of fragments should be taken into consideration:
$(1^{2k+1},0)_r$, $(1^{2k+1},0)_{\rho}$, $(1^{2k},0)_r$, $(1^{2k},0)_{\rho}$.
We denote their total numbers by $N_{2k+1}^{(r)}, N_{2k+1}^{(\rho)},
N_{2k}^{(r)}, N_{2k}^{(\rho)}$, respectively, and the corresponding ``probabilities'' by
$w_{2k+1}^{(r)}, w_{2k+1}^{(\rho)},w_{2k}^{(r)}, w_{2k}^{(\rho)}$, e.g.
$w_{2k+1}^{(r)}=N_{2k+1}^{(r)}/{\cal N}$. Here ${\cal N}$ is the total
number of sites of both types, $r$ and $\rho$. 
As in the prior consideration, a spin-0 state is attached from the right
to any finite spin-1 chain fragment.
Indices $r$ and $\rho$ specify the type of the rightmost site,
which is occupied by spin-0. 
Certainly, this definition incorporates
``empty'' chain fragments $0_r$ and $0_{\rho}$ into the scheme.
Not all the numbers $N$ mentioned above are independent, e.g., the leftmost 
sites of $(1^{2k},0)_r$ and $(1^{2k+1},0)_{\rho}$ are both of the $r$-type. 
They evidently follow all those chain fragments, which have a rightmost site of 
type $\rho$.
Thus, the following ``conservation law'' holds:
$$\sum_k(N_{2k}^{(r)}+N_{2k+1}^{(\rho)})=\sum_mN_m^{(\rho)},$$
which yields
\begin{equation}
\sum_kw_{2k}^{(r)}=\sum_kw_{2k}^{(\rho)}.
\label{n=n}
\end{equation}
The total number of lattice sites ${\cal N}$ expressed in terms of $N^{(r)}$
and $N^{(\rho)}$ gives rise to the equation:
\begin{equation}
1=\sum_{k\geq 0}(2(2k+1)w_{2k}^{(r)}+
(2k+2)(w_{2k+1}^{(\rho)}+w_{2k+1}^{(r)}),
\label{tot_num}
\end{equation}
and the energy per site is
\begin{eqnarray}
{\cal E}=\sum_k(2(\varepsilon_{2k}-k{\cal J}_0'-k{\cal J}_0)w_{2k}^{(r)}
\nonumber\\
+(\varepsilon_{2k+1}-(k+1){\cal J}_0'-k{\cal J}_0)w_{2k+1}^{(r)}
\nonumber\\
+(\varepsilon_{2k+1}-k{\cal J}_0'-(k+1){\cal J}_0)w_{2k+1}^{(\rho)}.
\label{tot_en}
\end{eqnarray}
In this equation we should use energies $\varepsilon_m$ given in Table 2.
Note, that the set $\{\varepsilon_{2k+1}\}$
coincides with $\{\epsilon_k\}$ used in model A. 

As we have assumed ${\cal J}_0'\geq {\cal J}_0$,
the contribution of chain fragments $(1^{2k+1},0)_{\rho}$ to the energy 
is not competative with the one for the $(1^{2k+1},0)_r$ fragments,
see Eq. (\ref{tot_en}). Thus, in the problem of finding the ground
state energy, we must check the contribution of two sorts of vertices: 
The first kind is defined as $w_{2k}^{(r)}=w_{2k}^{(\rho)}=1/(4k+2)$ 
with energy 
\begin{equation}
{\cal E}_{2k}=\frac 1{2k+1}(\varepsilon_{2k}-k{\cal J}_0'-k{\cal J}_0).
\label{even}
\end{equation}
For the second kind we get $w_{2k+1}^{(r)}=1/(2k+2)$ and
the energy reads
\begin{equation}
{\cal E}_{2k+1}=\frac 1{2k+2}
(\varepsilon_{2k+1}-(k+1){\cal J}_0'-k{\cal J}_0).
\label{odd}
\end{equation}
We shall denote corresponding regular structures as 
$\langle\langle  2k\rangle\rangle $ 
and $\langle\langle  2k+1\rangle\rangle $ in accordance with the numbers of 
spin-1 sites in chain fragments.
Note that in the second case the periodicity is $2k+2$, whereas in the first 
case it is $4k+2$.
Expressions (\ref{even})-(\ref{odd}) can be rewritten as 
\begin{equation}
{\cal E}_m=\frac 1{m+1}\left(\varepsilon_m-
m\frac{{\cal J}_0'+{\cal J}_0}2
-\theta_m\frac{{\cal J}_0'-{\cal J}_0}2\right),
\label{unif}
\end{equation}
where $\theta_m=0$ or 1 for $m$ even or odd. Except for the last term in the 
r.h.s., Eq.(\ref{unif}) has a form similar to Eq.(\ref{g.s.ener}). 
\vspace{-1.5cm}

\begin{figure}
\epsfxsize=80mm
\epsffile{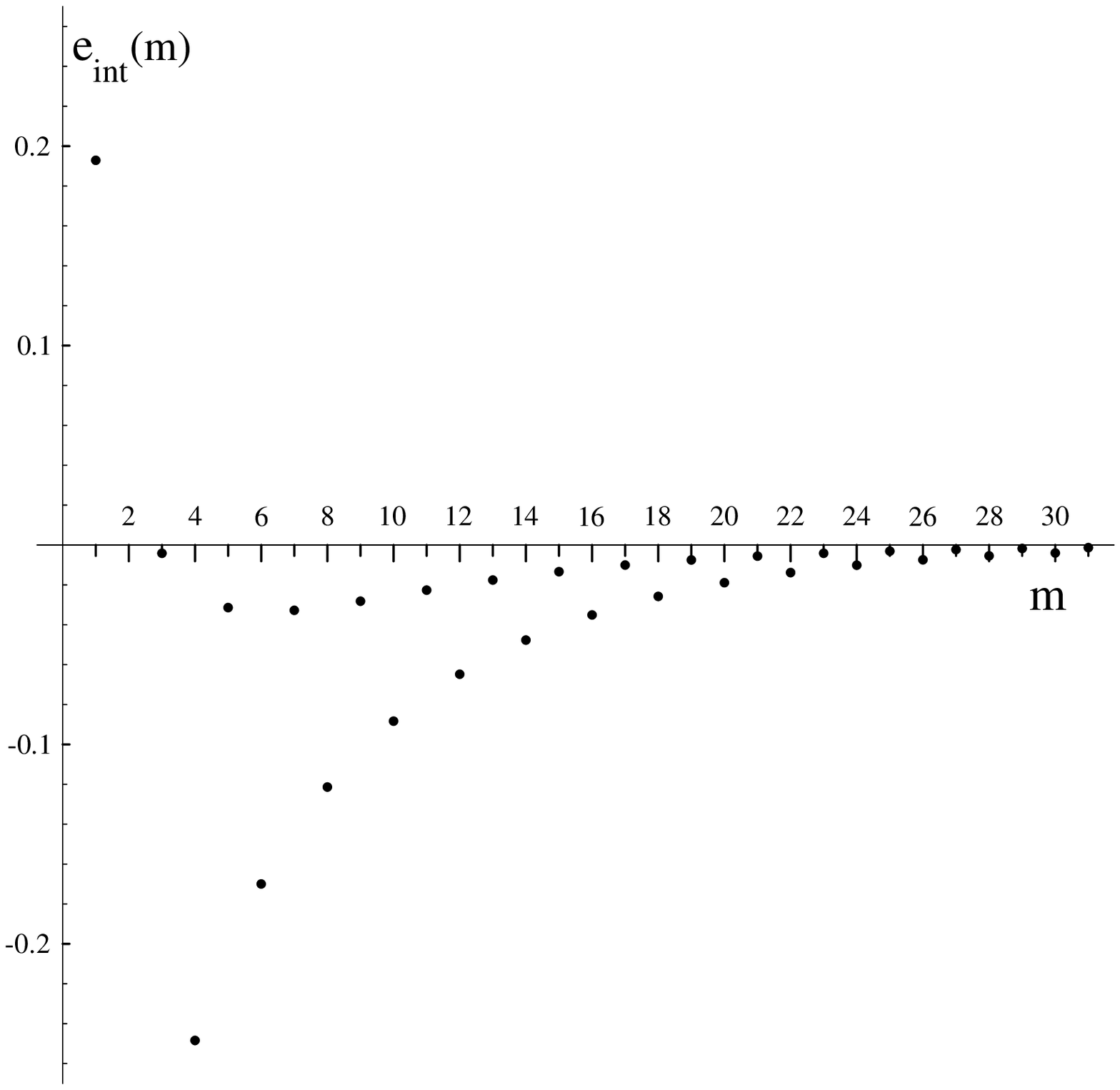}
\vspace{-1cm}

\noindent
\label{figure3}
\baselineskip=10pt
{{\small FIG.\ 3\quad  Shown are two subsets of $e_{\rm int}(m)$ for $m$
{\it odd} and {\it even}. The points for $m=2$ and $m=0$ are situated far 
below at -0.4056 and -1.2086, respectively. $e_{\rm int}(2k+1)$ of model B
coincides with $e_{\rm int}(k)$ of model A.}}
\end{figure}
\mbox{}
\vspace{-3mm}

\noindent
Shown in Fig. 3 is $e_{\rm int}$ at ${\cal J}_0'={\cal J}_0$. If, however, 
${\cal J}_0'\!>\!{\cal J}_0$, then the subset $\{e_{\rm int}(m,{\rm even})\}$ 
increases as compared with $\{e_{\rm int}(m,{\rm odd})\}$. Having Fig. 3 as a
prerequisite, we can describe all possible transformations as a function
of (${\cal J}_0'-{\cal J}_0$): 
\begin{itemize}
\item
As far as $e_{\rm int}(0)$ remains below $e_{\rm int}(7)$, which is the absolute
minimum of the $\{e_{\rm int}(m,{\rm odd})\}$ subset, our system only undergoes
the $\langle\langle  0 \rangle\rangle \rightarrow \langle\langle \infty\rangle\rangle $
transition;
\item 
If $e_{\rm int}(0)$ raises above $e_{\rm int}(7)$, but remains below the continuation
of the straight line
connecting the points at $m=5$ and at $m=7$,
then two transitions $\langle\langle  0 \rangle\rangle \rightarrow \langle\langle 7
\rangle\rangle \rightarrow \langle\langle \infty\rangle\rangle $ take place;
\item
Next, if $e_{\rm int}(0)$ raises above the continuation of the line connecting the points
at $m=5$ and $m=7$, but remains below the continuation of the line connecting the points
at $m=3$ and $m=5$, then the three transitions 
$\langle\langle  0 \rangle\rangle \rightarrow \langle\langle  5 \rangle\rangle \rightarrow
 \langle\langle  7 \rangle\rangle \rightarrow \langle\langle \infty\rangle\rangle $
take place.
\item
The scheme can be continued.
\item
Finally, at sufficiently large ${\cal J}_0'-{\cal J}_0$, we get a maximal 
possible number of first order transitions:
$\langle\langle  0 \rangle\rangle \rightarrow \langle\langle  1 \rangle\rangle \rightarrow
 \langle\langle  3 \rangle\rangle \rightarrow \langle\langle  5 \rangle\rangle \rightarrow
 \langle\langle  7 \rangle\rangle \rightarrow \langle\langle \infty\rangle\rangle $.
\end{itemize}
The phase diagram is shown in Fig. 4.
\vspace{-1.0cm}

\begin{figure}
\epsfxsize=80mm
\epsffile{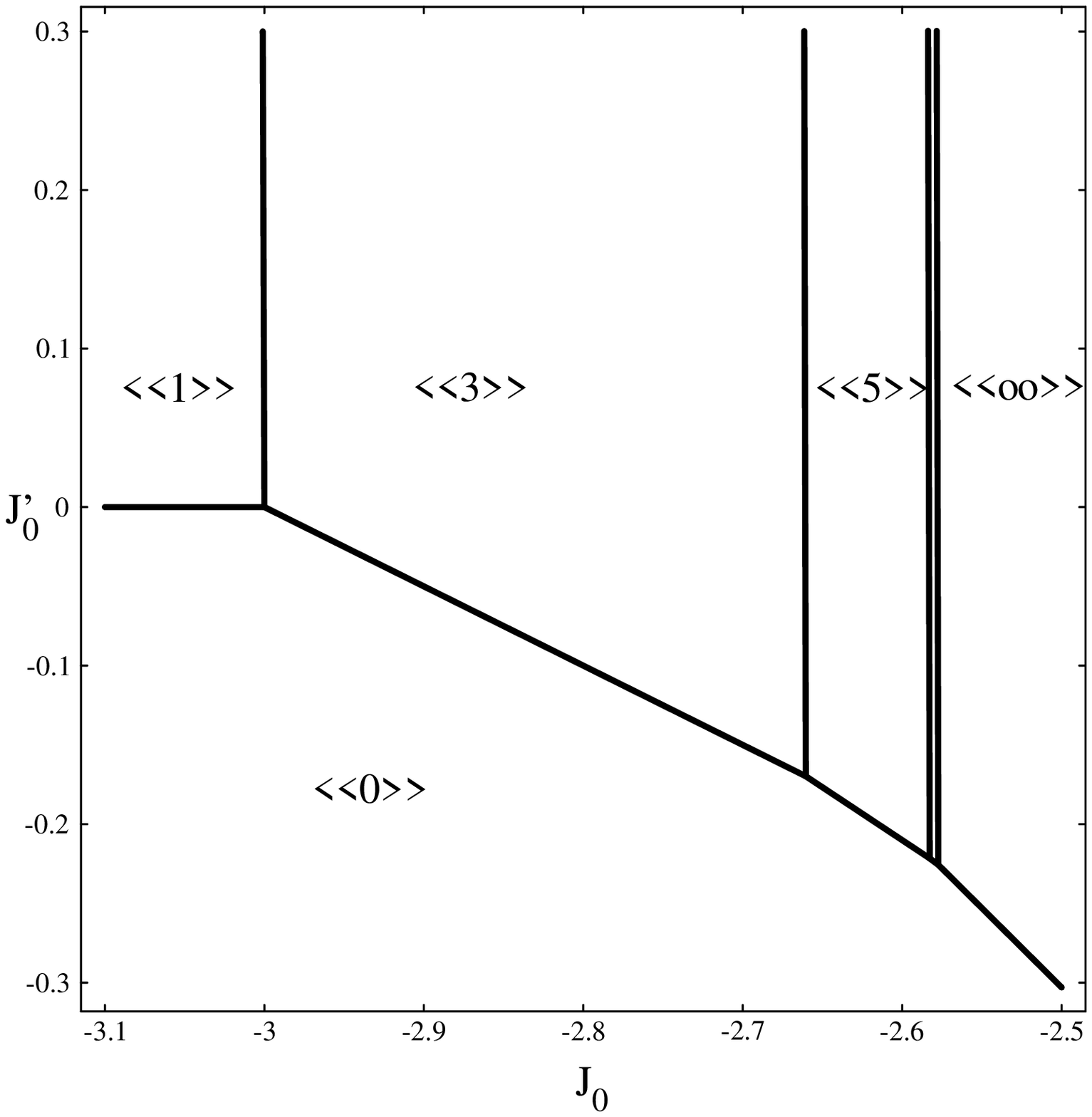}
\vspace{-1.5cm}

\noindent
\label{figure5}
\baselineskip=10pt
{{\small FIG.\ 4\quad  The phase diagram for model B is shown in the
(${\cal J}_0,{\cal J}_0'$) plane. A narrow area between $\langle\langle  5 \rangle\rangle $
and $\langle\langle \infty\rangle\rangle $ belongs to the $\langle\langle  7 \rangle\rangle $ phase. The line,
which separates the $\langle\langle  0 \rangle\rangle $-phase from the others, changes its slope from
$-1$ through $-3/4$, $-2/3$, $-1/2$ to 0.
}}
\end{figure}
\subsection*{Discussion and Conclusions}

The model considered in this paper is probably a simplest representative
of the family of Heisenberg models which possess an intrinsic property of
a hidden Ising symmetry. Compound spins of our model form a regular sublattice
within a one-dimensional chain: We have considered only the case of alternating 
chains, however, any periodicity within a system of compound spins is allowed.

We have restricted ourselves to the compound spin values 0 and 1. In principle,
a more complex construction for compound spins can be used, {\it i.e.}
instead of dumbbells three spins 1/2 may form a triangle. Then a compound 
spin is allowed to be 3/2 and 1/2. A treatment of this case is more
difficult, because all possible modulations in a distribution of spins
3/2 and 1/2 should be considered within the ensemble of infinite chains,
there are no spin-0 breakers like in the models of this work.
Another possibility would be associated with $\sigma$ constituents
of compound spins of a higher value than 1/2, say 1.  
Then, increasing ${\cal J}_0$ from large negative
values to moderate negative ones, we could pass through a few regimes,
starting from a periodic structure of elements $(s,0)$ through a few structures
whose periodicity elements are ($(s,1)^k,s,0$) to $(s,1)^\infty$, then, most
likely, a few intermediate structures $(s,2)^k,s,1$ will finally lead to a
perfect $(s,2)^\infty$ chain. However, the analysis of the phase diagram 
between perfect $(s,1)^\infty$ and $(s,2)^\infty$ structures is somewhat 
difficult because of the absence of spin-0 breakers.

For dealing with these more complicated systems, one could use 
a spin-wave approach, although tedious for finite systems, 
but well-defined. We have employed
this approach in order to compare our ``exact'' numerical results with
this approximate analytical scheme. For $s=1/2$ these two schemes qualitatively
lead to the same global minimum of $e_{\rm int}(k)$ at $k=1$, thus exhibiting 
the identical succession of transitions.
We have observed that for $s\geq 3/2$ $e_{\rm int}(k)$ is
{\it always} a monotonically decreasing function. This means that the occurence
of a negative minimum in $e_{\rm int}(k)$ should be due to non-linear terms of
a spin-wave expansion. It is clearly seen from Table 1 that this minimum
is extremely small ($\sim 10^{-4}e_{\rm int}(0)$) even at $s=3/2$ or 2,
but it exists and has an important influence on the succession of phase 
transitions vs ${\cal J}_0$. As this minimum occurs at $k=4$, which is a
rather short length, and thus is well-controlled numerically, we are absolutely sure 
of the existence of the minima. The data obtained from the spin-wave approach
are given in Table 3. Evidently they do not differ much from
our ``exact'' numerical results, decreasing very rapidly with $k$.
Exceptional is the $s=1$ case, where $e_{\rm int}(k)$ decreases
\begin{table}[b]
\squeezetable
\begin{tabular} {c|lr|lr|lr}
\multicolumn{1}{c|}{\mbox{}} & \multicolumn{2}{c|}{$s=1/2$} &
\multicolumn{2}{c|}{$s=3/2$} & \multicolumn{2}{c}{$s=2$}\\
\hline
$k$ &  $\epsilon_k$ & $e_{\rm int}(k)$ &  $\epsilon_k$ & $e_{\rm int}(k)$ 
& $\epsilon_k$ & $e_{\rm int}(k)$\\
\hline
0 & 0 &    0.370671 &      0 & 0.184008 &      0 & 0.131569 \\
1 & -2  & -0.192873 &     -4 & 0.012081 & -5 & 0.004480 \\
2 & -3.2928932  & -0.049311 & -7.8377223 & 0.002433 & -9.8768944 & 0.000497 \\
3 & -4.6937567  & -0.013719 & -11.667585 & 0.000644 & -14.750229 & 0.000073 \\
4 & -6.1203597  & -0.003866 & -15.496108 & 0.000194 & -19.623201 & 0.000012 \\
5 & -7.5540280  & -0.001079 & -19.324313 & 0.000063 & -24.496123 & 0.000002 \\
6 & -8.9897014  & -0.000297 & -23.152429 & 0.000022 & -29.369036 & 0.000001 \\
7 & -10.425941  & -0.000081 & -26.980516 & 0.000008 & -34.241947 & 0.000000 \\
8 & -11.862337  & -0.000022 & -30.808595 & 0.000003 & -39.114859 & 0.000000 \\
9 & -13.298777  & -0.000006 & -34.636671 & 0.000001 & -43.987770 & 0.000000 \\
10 & -14.735228  & -0.000002 & -38.464745 & 0.000000 & -48.860681  & 0.000000
\end{tabular}
\vspace{0.2cm}

\label{table3}
\baselineskip=10pt
{{\small TABLE\ 3\quad  Spin-wave approximation: 
The ground state energies of Hamiltonian $H_1$ and the function
$e_{\rm int}(k)$
for the chain fragments, consisting of $k$ spins 1 and $k+1$ spins $s$,
1/2, 3/2, and 2, are shown in three successive columns.}}
\end{table}
\mbox{}
\vspace{-0.4cm}

\noindent 
 slowly, as
within the ``exact'' numerical scheme, although monotonically.

We suppose that this important property of $e_{\rm int}(k)$ to exhibit
a minimum, more and more shallow, will be
valid for larger $s$-values, too, 
but we cannot say definitely whether the minimum remains at finite $k$-values
or shifts to $k\to\infty$. 

The next article in this series which will be published elsewhere
is devoted to the thermodynamics of mixed 
Heisenberg chains. Certainly, all the transitions will be smeared out due
to thermal fluctuations. However, the system must show big changes in
physical properties, such as the specific heat and the magnetic susceptibility,
if ${\cal J}_0$ is in a transitional area. In fact, as the chain
fragments of finite lengths possess total nonzero spins, which do not interact 
with spins of the nearest chain fragments, they form a system of paramagnetic
spins. Thus, the susceptibility will exhibit a Curie-like behaviour at low
temperatures. A prefactor will be temperature dependent, too, because the 
values of paramagnetic spins and their concentrations strongly correlate with the length
distribution of the chain fragments.

We illustrate
this by taking an example from the forthcoming article. Let us consider model A
with $s=1/2$ and ${\cal J}_0\approx -2$, {\it i.e.} where the chain fragments
$(s,1,s,0)$ and $(s,0)$ are only competitive. The former is practically
in a singlet state, the latter represents a purely paramagnetic spin 1/2,
their concentration in a lattice varies with $T$ as
$$
\frac 12\frac{1+\sqrt{1+4w}}{1+4w+\sqrt{1+4w}}, \quad w=\exp({\cal J}_0+2)/T.
$$
\subsection*{Acknowledgement}

This work has been performed within the research program of the Sonderforschungsbereich
341, K\"oln-Aachen-J\"ulich. We thank T. Kennedy for sending us some
unpublished results of numerical computations for spin-1 Heisenberg chains.
Discussions with A. Kl\"umper and A. Schadschneider 
were always informative to us.
\subsection*{Appendix A.}
\vspace{-2mm}

\renewcommand{\theequation}{A.\arabic{equation}}
\setcounter{equation}{0}
\begin{appendix}
To calculate the ground state energies of finite
chain fragments $(S_1,S_2)^k S_1$ we used a slightly modified
version of the well-known density matrix renormalization
group (DMRG) method introduced by White
\cite{white1,white2}.
Here we give a brief sketch of this method.

Let us consider the following finite fragment, which consists
of 3 spin-$S_1$ sites and 2 spin-$S_2$ sites:
\renewcommand{\arraystretch}{1.3}
\begin{eqnarray}
\label{frm_5chain}
\begin{array}{ccccc}
\fbox{$S_1$} & \fbox{$S_2$} & \fbox{$S_1$} & \fbox{$S_2$} & \fbox{$S_1$} \\
i_1          & i_2          & i_3          & i_4          & i_5
\end{array} \quad .
\end{eqnarray}
Each box denotes a single-spin Hilbert space with basis states numbered
by the corresponding index $i_\nu$ in the lower row. For the moment
$i_1,\,i_3,$ and $i_5$ are identical, as well as $i_2$ and $i_4$.
The Hamiltonian for the chain (\ref{frm_5chain}) can be divided into the
following contributions:
\begin{eqnarray}
\label{frm_5ham}
H=H^{B}_{i_1,j_1} \!+ H^{B}_{i_5,j_5} \!+ 
    H^{S_2}_{i_2,j_2} \!+ H^{S_2}_{i_4,j_4} \!+
    H^{S_1}_{i_3,j_3}\nonumber\\
 +\,H^{B S_2}_{i_1 i_2,j_1 j_2} \!+ H^{B S_2}_{i_5 i_4,j_5 j_4} \!+ 
    H^{S_1 S_2}_{i_3 i_2,j_3 j_2} \!+ H^{S_1 S_2}_{i_3 i_4,j_3 j_4}.
\end{eqnarray}
The first five terms are on-site contributions, the others couple
neighbouring sites. At this stage, $H^{B},\,H^{B S_2}$ are identical
to $H^{S_1},\,H^{S_1 S_2}$, respectively.

The first step of the algorithm now is to compute the lowest eigenvalue of
the Hamiltonian (\ref{frm_5ham}) and the corresponding eigenstate
by using the Lanczos method. The eigenvalue is directly the ground state
energy $\epsilon_2$ of the fragment $(S_1,S_2)^2 S_1$ and the corresponding
eigenstate $\phi_{i_1 i_2 i_3 i_4 i_5}$ serves as the `target state' of the
subsequent DMRG step. From this target state we calculate the density
matrix for the combined $i_1 i_2$ Hilbert space
\begin{equation}
\rho_{i_1 i_2,j_1 j_2} = \sum_{i_3,i_4,i_5} \phi_{i_1 i_2 i_3 i_4 i_5}
                                            \phi_{j_1 j_2 i_3 i_4 i_5} \,.
\end{equation}
Clearly, the eigenvectors of $\rho$ to the largest eigenvalues give the
most important contributions to the target state
$\phi_{i_1 i_2 i_3 i_4 i_5}$. The idea of the DMRG method is to reduce
the dimension of the $i_1 i_2$ Hilbert space by projecting all operators
onto the eigenstates of $\rho$ belonging to the $N$ largest eigenvalues. Let $U$
be the rectangular matrix that contains these $N$ normalized eigenvectors as
columns\cite{identity}. This matrix $U$ allows
us to combine the $i_1$ and $i_2$ Hilbert spaces into a new single Hilbert space
$i'_1$ while limiting its dimension to at most $N$. All operators $A$ acting on
$i_1 i_2$ are transformed into operators $A'$ acting on $i'_1$ via
\begin{equation}
A'=UAU^\dagger \,.
\end{equation}
We then arrive at the situation
\begin{equation}
\begin{array}{ccc}
\fbox{$S_1\; S_2$} & \fbox{$S_1$} & \fbox{$S_2\; S_1$} \\
i'_1               & i_3          & i'_5
\end{array} \quad .
\end{equation}
Note that $i'_5$ is the reflected version of $i'_1$.
The length of the chain can now be increased by inserting an $(S_1,\,S_2)$ pair,
yielding
\begin{equation}
\label{frm_kchain}
\begin{array}{ccccc}
\fbox{$S_1\; S_2$} & \fbox{$S_1$} & \fbox{$S_2$} & \fbox{$S_1$} & \fbox{$S_2\; S_1$} \\
i'_1               & i'_2         & i'_3         & i'_4         & i'_5
\end{array} \quad ,
\end{equation}
where we have renamed $i_3$ to $i'_2$.
The Hamiltonian for the chain (\ref{frm_kchain}) again has the structure
(\ref{frm_5ham}), but with interchanged roles of $S_1$ and $S_2$ and new
operators
\begin{equation}
\label{frm_kham}
\begin{array}{rcl}
H^{B}_{\mathrm{new}} &=& U \left[ H^{B}\otimes 1   +
                                  1\otimes H^{S_2} +
                                  H^{B S_2} \right] U^\dagger \\
H^{B S_1}_{\mathrm{new}} &=&
  \sum_{\nu} \left(U \left[ 1\otimes B^\nu \right] U^\dagger\right)\otimes A^\nu \,,
\end{array}
\end{equation}
provided the original interaction is given by
\begin{equation}
H^{S_1 S_2} = \sum_{\nu} A^\nu \otimes B^{\nu} \; .
\end{equation}
This completes the DMRG step. The Lanczos method is now applied to the new
full Hamiltonian for chain (\ref{frm_kchain}), the lowest eigenvalue gives
the energy $\epsilon_3$ of the chain fragment $(S_1,\,S_2)^3 S_1$ and the corresponding
eigenvector serves as the target state for the subsequent DMRG step.

The algorithm described above is a straightforward generalization of the
standard `infinite system method' of White\cite{white1} to alternating
symmetric spin chains. It differs from the original algorithm in two points:
1. It contains an additional spin $i_3$ in the center of the chain, and
2. the role of $S_1$ and $S_2$ has to be interchanged at every DMRG step.
We did not use the
extended `finite system method', as agreement with competing pure Lanczos
calculations was already satisfactory and could be systematically increased
by using larger $N$.
\end{appendix}
\subsection*{Appendix B.}
\vspace{-2mm}

\renewcommand{\theequation}{B.\arabic{equation}}
\setcounter{equation}{0}
\begin{appendix}
In this Appendix we show how a succession of first order transitions takes place.
We start by noting that the $\langle 0\rangle $
structure is optimal when ${\cal J}_0$ has a large negative value and
the $\langle \infty\rangle $-phase is realized at large positive ${\cal J}_0$.

Let us define the ${\cal J}_0$ parameter as ${\cal J}_0^{(m,n)}$,
when two phases, $\langle m\rangle $ and $\langle n\rangle $, are at equilibrium, {\it i.e.},
${\cal E}_m={\cal E}_n$.
Then we can employ representation (\ref{ener_fit}) to determine the following
equations:
\begin{eqnarray}
e_{\infty}-{\cal J}_0^{(k-1,k)}-e_0 \nonumber\\
=(k+1)e_{\rm int}(k-1)-ke_{\rm int}(k),\label{j_nn1}
\\
e_{\infty}-{\cal J}_0^{(k,k+1)}-e_0 \nonumber\\
=(k+2)e_{\rm int}(k)-(k+1)e_{\rm int}(k+1).
\label{j_nn2}
\end{eqnarray}
From Eqs.(\ref{j_nn1})-(\ref{j_nn2}) we obtain
\begin{eqnarray}
{\cal J}_0^{(k,k+1)}-{\cal J}_0^{(k-1,k)} \nonumber\\
=(k+1)(e_{\rm int}(k+1)-2e_{\rm int}(k)+
e_{\rm int}(k-1)).\label{jj_nn}
\end{eqnarray}

Thus, we come to the {\it first} conclusion: If $e_{\rm int}(k)$ is concave upwards
at any $k$, we have a full succession of transitions:
$\langle 0\rangle \rightarrow\langle 1\rangle \rightarrow\langle 2\rangle \rightarrow
 \ldots \rightarrow\langle \infty\rangle$.

Suppose now that the condition for $e_{\rm int}(k)$ does not hold.
For instance, if for $0<j<m$ ~$e_{\rm int}(k+j)$ satisfies inequalities:
\begin{eqnarray}
e_{\rm int}(k+j)>e_{\rm int}(k)+\frac jm(e_{\rm int}(k+m)-e_{\rm int}(k)),
\label{ineq}
\end{eqnarray}
then a generalization of Eqs. (\ref{j_nn1})-(\ref{j_nn2}) is
\begin{eqnarray}
{\cal J}_0^{(k,k+j)}=e_{\infty}-e_0-e_{\rm int}(k) \nonumber\\
+\frac{k+1}j(e_{\rm int}(k+j)-e_{\rm int}(k)).
\label{j_nn3}
\end{eqnarray}
Taking into account inequalities (\ref{ineq}), one simply obtains
${\cal J}_0^{(k,k+j)}>{\cal J}_0^{(k,k+m)}$. These last inequalities
show that all intermediate phases whose ``energies'' $e_{\rm int}(\ell)$
are above the enveloping line, cannot be realized as ground states at any
value of ${\cal J}_0$.
An important consequence is that if $e_{\rm int}$ achieves an absolute 
minimum at some $k$, this results in a direct transition
$\langle k\rangle \rightarrow \langle \infty\rangle $.
A convenient expression for the critical value ${\cal J}_0^{(k,\infty)}$
can be derived from (\ref{j_nn3}):
\begin{eqnarray}
{\cal J}_0^{(k,\infty)}\!={\cal J}_0^{(k-1,k)}\!+
(k+1)(e_{\rm int}(k-1)-e_{\rm int}(k)).
\label{infty}
\end{eqnarray}
\end{appendix}


\begin{references}
\bibitem[*]{gu} On leave from Landau Institute for Theoretical Physics,
Chernogolovka, Moscow Region 142432, Russia.
\bibitem{vega} H.-J.~de Vega and F.~Woynarovich, J. Phys. A {\bf 25}, 4499 
(1992).
\bibitem{fujii} M. Fujii, S. Fujimoto and N. Kawakami, J. Phys. Soc. Jap. 
{\bf 65}, 2381 (1996).
\bibitem{doerfel}  B.-D.\ D\"orfel and S.\ Mei\ss ner, 
J. Phys. A {\bf 30}, 1831 (1997).
\bibitem{pati}     S.K. Pati, S. Ramasesha and D. Sen, e-print cond-mat/9610080
\bibitem{brehmer}  S. Brehmer, H.-J. Mikeska and S. Yamamoto, e-print cond-mat/9610109
\bibitem{kolezhuk} A.K. Kolezhuk, H.-J. Mikeska and S. Yamamoto, e-print cond-mat/9610097
\bibitem{alcaraz} F.C. Alcaraz and A.L. Malvezzi, e-print cond-mat/9611227,
to appear in J.~Phys.~A: Math.~Gen
\bibitem{white1} S.R. White, Phys. Rev. B {\bf 48}, 10345 (1993).
\bibitem{white2} S.R. White, Phys. Rev. Lett. {\bf 69}, 2863 (1992);
S.R. White and D.A. Huse, Phys. Rev. B {\bf 48}, 3844 (1993).
\bibitem{hubb} J. Hubbard, Phys. Rev. B {\bf 17}, 494 (1978).
\bibitem{u_p1} V.L. Pokrovsky and G.V. Uimin, J.Phys. C: Solid State Phys. 
{\bf 11}, 3535 (1978).
\bibitem{f_s} M.E. Fisher and W. Selke, Phys. Rev. Lett. {\bf 44}, 1502 (1980);
M.E. Fisher and W. Selke, Phil. Trans. {\bf 302}, 1 (1981).
\bibitem{u_p2} V.L. Pokrovsky and G.V. Uimin, J. Phys. C: Solid State Phys.
{\bf 15}, L 353 (1982); 
V.L. Pokrovsky and G.V. Uimin, Sov. Phys.-JETP {\bf 55}, 950 (1982).
\bibitem{kennedy} T. Kennedy, J. Phys.: Condens.Matter {\bf 2}, 5737 (1990).
\bibitem{comment} Numerical computations show that $e_0$ has
the same value for chain fragments of even and odd lengths.
\bibitem{identity} If the dimension of the combined $i_1 i_2$ Hilbert space
is less than $N$, we define $U$ to be the identity matrix.
\end{references}
\end{document}